\documentclass[10pt]{article}
\usepackage{fullpage}
\usepackage{amsmath}
\usepackage{amssymb}
\usepackage{amsfonts}
\usepackage[dvips]{epsfig}
\usepackage{color}

\def\##1{\underline{#1}}
\def\=#1{\underline{\underline{#1}}}

\def\+#1{\underline{\bf #1}}
\def\*#1{\underline{\underline{\bf #1}}}

\def\r#1{(\ref{#1})}
\def\l#1{\label{#1}}
\def\c#1{\cite{#1}}

\def\le{\left(}
\def\ri{\right)}
\def\les{\left[}
\def\ris{\right]}
\def\lec{\left\{}
\def\ric{\right\}}

\def\.{\mbox{ \tiny{$^\bullet$} }}

\def\eps{\varepsilon}

\def\muo{\mu_{\scriptscriptstyle 0}}

\begin{document}

\begin{center}

\LARGE{ {\bf Voigt-wave propagation in active  materials
}}
\end{center}
\begin{center}
\vspace{10mm} \large

 Tom G. Mackay\footnote{E--mail: T.Mackay@ed.ac.uk.}\\
{\em School of Mathematics and
   Maxwell Institute for Mathematical Sciences\\
University of Edinburgh, Edinburgh EH9 3FD, UK}\\
and\\
 {\em NanoMM~---~Nanoengineered Metamaterials Group\\ Department of Engineering Science and Mechanics\\
Pennsylvania State University, University Park, PA 16802--6812,
USA}\\
 \vspace{3mm}
 Akhlesh  Lakhtakia\footnote{E--mail: akhlesh@psu.edu}\\
 {\em NanoMM~---~Nanoengineered Metamaterials Group\\ Department of Engineering Science and Mechanics\\
Pennsylvania State University, University Park, PA 16802--6812, USA}

\normalsize

\end{center}

\begin{center}
\vspace{15mm} {\bf Abstract}

\end{center}
If a dissipative anisotropic dielectric material, characterized by the permittivity matrix $\=\eps$, supports Voigt-wave propagation, then so too does the analogous active material characterized by the permittivity matrix $\={\tilde{\eps}}$, where $\={\tilde{\eps}}$ is the hermitian conjugate of $\=\eps$. Consequently, a dissipative material that supports Voigt-wave propagation can give rise to a material that supports the propagation of Voigt waves with attendant linear gain in amplitude with propagation distance, by infiltration with an active dye.

\vspace{14mm}

\noindent {\bf Keywords:} Voigt-wave propagation, active materials, anisotropic dielectric materials

\vspace{14mm}

Plane-wave analysis is a cornerstone in the electromagnetic theory.
Although
plane waves themselves are idealizations, possessing limitless spatial and
temporal extents and limitless energy,
they can facilitate deep  insights
 into fields distant from
sources and, by means of superpositions of plane waves,  realistic signals  may be conveniently represented.
Therefore, undergraduate textbooks on electromagnetics and optics invariably present plane waves right after introducing the Maxwell equations and the standard boundary conditions \cite{Hayt,Iskander,Ida}.

In nondissipative dielectric materials (and vacuum),  plane waves are straightforwardly characterized.
 These waves propagate with only one phase speed, with arbitrary states of polarization, in any direction. Accordingly, such materials are called unirefringent. In nondissipative anisotropic dielectric materials, generally two distinct plane waves can propagate in any direction, with their phase speeds and polarization states being dependent upon the
   propagation  direction \c{BW,Chen}. Accordingly, such materials are called birefringent.
In any uniaxial dielectric material, ordinary plane waves propagate with a phase speed that is independent of the propagation direction whereas extraordinary plane waves propagate with a  phase speed that depends upon the propagation direction. Along the two directions aligned with the optic axis of the material, the phase speeds of the ordinary and the extraordinary plane waves are the same. In  a biaxial dielectric material, generally two extraordinary plane waves can propagate, at two different phase speeds, depending upon propagation direction. There are two optic axes for these materials, with the phase speeds of both plane waves being the same for propagation along an optic axis.

 Matters are further complicated if the materials are dissipative. In certain dissipative anisotropic dielectric materials, there  exist propagation directions~---~distinct from the directions aligned with the optic axes~---~along which plane waves have  only one phase speed. A plane wave that propagates along such a direction is called a Voigt wave \c{Voigt,Panch,Khap,Fedorov,Ranganath}.
A Voigt wave represents a singular form of plane-wave propagation that can arise inside
certain
dissipative biaxial dielectric materials of the monoclinic and triclinic types \c{ML_PiO}, as well as inside certain gyrotropic materials \c{Agranovich,Grech},
but neither   inside  nondissipative materials nor inside isotropic or uniaxial dielectric materials.
 Mathematically, Voigt-wave propagation arises when an eigenvalue of the corresponding plane-wave propagation matrix  has an algebraic multiplicity which exceeds its geometric multiplicity \c{ML_PiO,TAW}. A crucial difference between a Voigt wave and a plane wave propagating along an optic axis is that the amplitude of the Voigt wave
varies linearly with propagation distance whereas this is not the case for the plane wave propagating along the optic axis.

While Voigt waves were first investigated experimentally and theoretically
for  pleochroic minerals \c{Voigt,Panch,Ranganath},
  greater scope for realizing Voigt-wave  propagation is offered by engineered materials \c{Lakh_helicoidal_bianisotropic_98,Berry}, especially homogenized composite
materials \c{ML03,ML_WRM}.  The directions in which Voigt waves propagate in a
  homogenized composite
material may be controlled through careful design of the microstructure and selection of the
 constitutive properties of the component materials \c{M2011_JOPA}, or by the application of an external field in the case of electro-optic materials \c{Voigt_Pockels}. Indeed,
 the ability to control the directions for Voigt wave propagation is promising
 for  technological applications such as optical sensing \c{M2014_JNP}.

All previous works on Voigt waves have focused on
 dissipative materials. The issue of Voigt waves in  active materials~---~that is, materials in which plane-wave propagation is accompanied by a net flow of energy  from the material to the field~---~has not been addressed hitherto.
 Active materials are useful, or indeed essential, for a host of  technoscientific applications. Prime examples are provided by
 lasing materials,
  scintillators, and  luminescent solar concentrators \cite{Hecht,Vij,Yukihara,Debije}. In this context, let us, for example, single out doped tungstate materials, which can both support Voigt-wave propagation courtesy of their  monoclinic crystal symmetry and function as  highly efficient lasing materials \c{JJAP,Physics_Procedia}.

 Analytical treatment of Voigt waves   involves electromagnetic theory, and a commensurate level of mathematics, accessible to senior undergraduate students of the physical sciences and related engineering disciplines. In particular, the following presentation highlights the usefulness of matrix algebra and complex numbers in physics, and introduces the notion of active materials which is largely conspicuous by its absence in undergraduate electromagnetics courses.

Let us consider a general anisotropic dielectric material,
   characterized by the frequency-domain constitutive relations \c{ML_PiO}
\begin{equation} \l{CR}
\left.
\begin{array}{l}
\#D (\#r) = \=\eps \cdot \#E(\#r) \vspace{4pt} \\
\#B (\#r) = \muo \#H(\#r)
\end{array}
\right\},
\end{equation}
where the components $\eps_{\ell j}$ ($\ell, j \in \lec 1,2,3 \ric$) of the permittivity matrix
\begin{equation}
\=\eps = \le \begin{array}{ccc}
\eps_{xx} & \eps_{xy} & \eps_{xz} \\
\eps_{yx} & \eps_{yy} & \eps_{yz} \\
\eps_{zx} & \eps_{zy} & \eps_{zz}
\end{array}
\ri
\end{equation}
are complex valued
and $\muo = 4 \pi \times 10^{-7}$ H m${}^{-1}$ is the permeability of free space. Without loss of generality, let a Voigt wave propagate  along the $z$ axis.
Then,
an eigenvalue/eigenvector analysis  reveals that the conditions for Voigt-wave propagation are met when the following two conditions are  satisfied \c{Gerardin}:
\begin{itemize}
\item[(i)] $\mathcal{C}_a (\=\eps) = \le \delta_{11} - \delta_{22} \ri^2 + 4 \delta_{12} \delta_{21} = 0 $, and
\item[(ii)] $\mathcal{C}_{b1} (\=\eps) = \delta_{12} \neq 0 $ and/or $\mathcal{C}_{b2} (\=\eps) = \delta_{21} \neq 0 $,
\end{itemize}
where the scalar parameters
\begin{equation}
\left.
\begin{array}{l}
\delta_{11} = \eps_{xx} \eps_{zz} - \eps_{xz} \eps_{zx} \vspace{4pt}
\\
\delta_{12} = \eps_{xy} \eps_{zz} - \eps_{xz} \eps_{zy} \vspace{4pt}
\\
\delta_{21} = \eps_{yx} \eps_{zz} - \eps_{yz} \eps_{zx} \vspace{4pt}
\\
\delta_{22} = \eps_{yy} \eps_{zz} - \eps_{yz} \eps_{zy}
\end{array}
\right\}.
\end{equation}

 The  flow of energy associated  with  propagation is represented by the time-averaged Poynting vector $\langle \#P (\#r) \rangle_t = \le 1/2 \ri \mbox{Re} \les \#E (\#r) \times \#H^*(\#r) \ris$, with the superscript ${}^*$ denoting the complex conjugate and the operator $\mbox{Re} \les \cdot \ris$ delivering the real part. In particular, the divergence of the
  time-averaged Poynting vector is given by \c{Chen}
  \begin{equation} \l{PV}
  \#\nabla \cdot \langle \#P (\#r) \rangle_t = \mbox{Re} \lec \frac{i \omega}{2} \les \, \#H^*(\#r) \cdot \#B (\#r) - \#E (\#r) \cdot \#D^* (\#r) \, \ris \ric,
  \end{equation}
  where $\omega > 0$ is the angular frequency.
For  the anisotropic dielectric material characterized by the constitutive relations \r{CR}, the divergence \r{PV} simplifies to \c{Chen}
\begin{equation} \l{gPV}
  \#\nabla \cdot \langle \#P (\#r) \rangle_t = \frac{i \omega}{4} \, \#E^*(\#r) \cdot \le \=\eps - \={\tilde{\eps}} \ri \cdot \#E (\#r),
  \end{equation}
where  $\={\tilde{\eps}}$ represents the hermitian conjugate of $\=\eps\,$; i.e.,  the components of $\={\tilde{\eps}}$ are $\tilde{\eps}_{\ell j} = \eps^*_{j \ell}$ ($\ell\in \lec x,y,z \ric$ and $ j \in \lec x,y,z \ric$).

Now,
  if the material characterized by the constitutive relations \r{CR} is:
   \begin{itemize}
\item   dissipative then $ \#\nabla \cdot \langle \#P (\#r) \rangle_t < 0$, which,  according to Eq.~\r{gPV},  implies that the three eigenvalues of the matrix $i  \le \=\eps - \={\tilde{\eps}} \ri$ are negative valued (i.e., the matrix $i  \le \=\eps - \={\tilde{\eps}} \ri$ is negative definite) \c{TAW};
  \item  active then $ \#\nabla \cdot \langle \#P (\#r) \rangle_t > 0$, which,  according to Eq.~\r{gPV}, implies that the three eigenvalues of the matrix  $i  \le \=\eps - \={\tilde{\eps}} \ri$ are positive valued (i.e., the matrix
      $i  \le \=\eps - \={\tilde{\eps}} \ri$ is positive definite) \c{TAW}.
 \end{itemize}
 Therefore, if the material possessing the permittivity matrix $\=\eps$ is dissipative then the material described by the permittivity matrix $\={\tilde{\eps}}$ is active, and vice versa.

 Let us turn to the conditions (i) and (ii) that must be satisfied for Voigt-wave propagation. After some algebraic manipulations, one finds
 \begin{equation}
 \left.
 \begin{array}{l}
 \mathcal{C}_a (\={\tilde{\eps}}) = \mathcal{C}^*_a (\=\eps)  \vspace{4pt} \\
 \mathcal{C}_{b1} (\={\tilde{\eps}}) = \mathcal{C}^*_{b2} (\=\eps)  \vspace{4pt} \\
 \mathcal{C}_{b2} (\={\tilde{\eps}}) =  \mathcal{C}^*_{b1} (\=\eps)  \end{array}
 \right\}.
 \end{equation}
Therefore, if a dissipative (or active)  material
described by the permittivity matrix $\=\eps$ supports Voigt-wave propagation then so too must the analogous active (or dissipative) material described by the permittivity matrix $\={\tilde{\eps}}$.

An important implication of this finding is as follows. Suppose we consider a certain dissipative material, say a porous homogenized composite material with monoclinic biaxial symmetry. Voigt waves can propagate in this material with  attendant linear attenuation as the propagation distance increases \c{ML03}. Now if this material were to be infiltrated with a sufficient quantity of  an active dye (which is fluorescent \cite{Jameson}), such that the monoclinic symmetry remained unchanged, then Voigt waves could propagate in this material with attendant linear gain as the propagation distance increased. Thus, 
one could linearly amplify signals propagating along a specific direction in addition to
the exponential amplification,
while along other directions the  signal amplification would be just exponential.

\vspace{15mm}
\noindent {\bf Acknowledgements.  } TGM acknowledges the support of EPSRC grant EP/M018075/1.
AL thanks the Charles Godfrey Binder Endowment at Penn State for ongoing support of his research activities.

\end{document}